\documentclass[aip,jcp,reprint,floatfix]{revtex4-2}

\usepackage{graphicx}
\usepackage[utf8]{inputenc}
\usepackage[T1]{fontenc}
\usepackage{amsmath}
\usepackage{bm}
\usepackage{amssymb}
\usepackage{amsmath}
\usepackage{float}
\usepackage{hyperref}

\begin{document}
\title{Self-Tuning Hamiltonian Monte Carlo for Accelerated Sampling}

\author{Henrik Christiansen}
\email{henrik.christiansen@neclab.eu}
\author{Federico Errica}%
\email{federico.errica@neclab.eu}
\author{Francesco Alesiani}%
\email{francesco.alesiani@neclab.eu}
\affiliation{NEC Laboratories Europe GmbH, Kurfürsten-Anlage 36, 69115 Heidelberg, Germany.}
\date{\today}

\begin{abstract}
    The performance of Hamiltonian Monte Carlo simulations crucially depends on both the integration timestep and the number of integration steps. 
    We present an adaptive general-purpose framework to automatically tune such parameters, based on a local loss function which promotes the fast exploration of phase-space.
    We show that a good correspondence between loss and autocorrelation time can be established, allowing for gradient-based optimization using a fully-differentiable set-up.
    The loss is constructed in such a way that it also allows for gradient-driven learning of a distribution over the number of integration steps.
    Our approach is demonstrated for the one-dimensional harmonic oscillator and alanine dipeptide, a small protein common as a test case for simulation methods.
    Through the application to the harmonic oscillator, we highlight the importance of not using a fixed timestep to avoid a rugged loss surface with many local minima, otherwise trapping the optimization.
    In the case of alanine dipeptide, by tuning the only free parameter of our loss definition, we find a good correspondence between it and the autocorrelation times, resulting in a $>100$ fold speed up in optimization of simulation parameters compared to a grid-search.
    For this system, we also extend the integrator to allow for atom-dependent timesteps, providing a further reduction of $25\%$ in autocorrelation times.
\end{abstract}

\maketitle

\section{Introduction}
Simulations of molecular systems are predominately performed using either molecular dynamics (MD)\cite{hollingsworth2018molecular} or (random walk) Monte Carlo (MC)\cite{vitalis2009methods} simulations.
While there are some peculiarities of these methods, both approaches can be used to sample the canonical ensemble, i.e., a system in contact with a heatbath, and can, for many purposes, be used as plug-in replacements for sampling.
\par
In the literature, there already exist a couple of approaches to combine both methods,\cite{frenkel2001understanding,allen2017computer} where one of the most studied ones is Hamiltonian Monte Carlo (HMC),\cite{neal2011mcmc,neal2012bayesian,betancourt2017conceptual} which was originally introduced as hybrid Monte Carlo.\cite{duane1987hybrid}
The basic idea is to propagate the system using a (microcanonical) integrator for a given number of steps, which conserves the total energy as used in MD simulations but adds an acceptance of proposals obtained in this way using MC.
In case of rejection of a move, the system is reset (in the typical MC fashion) to the previous state.
The additional step required for HMC is to randomly draw new velocities, as otherwise, the resulting configuration would be identical after rejection.
Compared to the individual methods on their own, this combined approach has the advantage that there is no adverse effect of the numerical integration or external control of temperature while providing a systematic way to propose trial configurations.
\par
The trial configurations proposed by this approach can have a high acceptance probability, especially for small timesteps and few integration steps, as then both the numeric error and the integration errors are small (the total energy is in principle conserved by a microcanonical integrator).
Choosing both these parameters small, however, leads to slow phase space exploration, whereas choosing them large results in the simulation of a shadow Hamiltonian and fast accumulation of numeric errors resulting in small acceptance rates.
This implies that there is some balance between those two effects, for which the simulation is much faster at exploring the phase space.
Sub-optimal values for these two parameters can severely limit the efficiency of the simulation, reflected in a large autocorrelation time.
\par
We introduce a fully-differentiable framework that allows to tune these simulation parameters of HMC via backpropagation\cite{rosenblatt1961principles,rumelhart1985learning} based on a local loss definition.
This, ultimately, replaces the otherwise needed expensive grid search for good simulation parameters.
The framework is applied to the one-dimensional harmonic oscillator, highlighting subtle effects of the approach otherwise hidden in larger systems, and alanine dipeptide,\cite{hermans2011amino} a paradigmatic system for novel simulation techniques.
For the harmonic oscillator, we focus on fundamental properties of HMC, in particular the connection between fixed timesteps and local minima in the loss.
For alanine dipeptide, we find that using the usual definition of loss from classical adaptive MC literature provides no good correspondence to the autocorrelation times of the potential energy.
We propose a loss with only one hyper-parameter and show that it can be tuned to provide a substantially improved correlation.
Further, we extended the integrator to include atom-dependent timesteps leading to additional acceleration, which would be very difficult to optimize for using heuristic (gradient uninformed) methods.
\par
The paper is organized as follows:
In Section~\ref{sec:sim_met} we give a short review of the involved simulation methods, followed by an introduction to our approach in Section~\ref{sec:STHMC}.
Section~\ref{sec:res} presents results for the two systems, 
Finally, we will conclude and give an outlook on future research in Section~\ref{sec:con}.

\section{Review of Simulation Methods}
\label{sec:sim_met}
In the following, our goal is to simulate a classical system, which consists of particles/atoms interacting via classical potentials.
This system is coupled to a heatbath, i.e., our target is to simulate in the canonical ensemble.
Then, each microstate described by the spatial coordinates $\bm{x}$ occurs with a probability that is given by 
\begin{equation}
    P^{\mathrm{eq}}(\bm{x})=\frac{1}{Z}e^{-U(\bm{x})/(k_BT)},
    \label{eq:peq}
\end{equation}
where $Z=\int e^{-U(\bm{x})/k_BT} d\bm{x}$ is the partition function (in statistics often simply referred to as normalizing constant), $U(\bm{x})$ is the (potential) energy of a microstate depending on the atoms' positions, $k_B$ is the Boltzmann constant, and $T$ is the temperature of the heatbath.
While the evaluation of the partition function would provide access to many thermodynamic observables, this is in practice not possible since this necessitates evaluation of all possible microstates.
Instead, one attempts to approximate expectation values of quantities of interest at a fixed temperature by producing samples from the target distribution, utilizing methods that do not rely on the value of $Z$.

\subsection{Monte Carlo}
The main idea behind an MC simulation is to build a Markov chain, starting with a random configuration of the system at interest and subsequently progressing by proposing new configurations (which only depend on the current configuration).\cite{newman1999monte,janke2013monte}
There is some freedom in choosing the transition probabilities $W_{kl}=W(\bm{\xi}_k,\bm{\xi}_l)$ between microstate $\bm{\xi}_k$ and $\bm{\xi}_l$.
In MC for molecular systems, typically $\bm{\xi}=\bm{x}$, i.e., a set of Cartesian coordinates, but in general this can be any configurational information.
One of the most flexible choices for the acceptance criterion between states is the original Metropolis algorithm\cite{metropolis1953equation} which reads
\begin{equation}
    w_{kl}=w(\bm{\xi}_k,\bm{\xi}_l)=\min\left(1,\frac{f(\bm{\xi}_k,\bm{\xi}_l)}{f(\bm{\xi}_k,\bm{\xi}_l)}\frac{P^{\mathrm{eq}}(\bm{\xi}_l)}{P^{\mathrm{eq}}(\bm{\xi}_k)}\right),
    \label{eq:metropolis}
\end{equation}
where $f_{kl}=f(\bm{\xi}_k,\bm{\xi}_l)$ is the proposal probability for a potential update to a new microstate.
Here, the partition function cancels, since one is only interested in the ratio of the equilibrium distributions.
This then leads to the transition probability
\begin{equation}
    W_{kl}=\begin{cases}
        f_{kl}w_{kl} & k \neq l \\
        f_{kl} + \sum_{k \neq l} f_{kl}(1-w_{kl}) & k=l
    \end{cases}.
\end{equation}
Using this prescription, it is easy to see that the detailed balance condition given by 
\begin{equation}
    W_{kl}P^{\mathrm{eq}}_k=W_{lk}P^{\mathrm{eq}}_l
\end{equation}
is fulfilled.
This is a sufficient condition for the convergence to the equilibrium distribution.
\par
There are many ways to propose configurational changes to the system, which then constitute the move set.
The ``optimal'' set depends highly on the system and its parameters, where the parameters are often optimized based on some target acceptance rate or a local criterion based on the movement in phase space, such as the expected squared jump distance.\cite{pasarica2010adaptively}
One example of a more systematic approach to tuning such parameters in a classical MC simulation is Ref.~\onlinecite{bojesen2018policy}, where parameters of distributions used to propose changes are optimized based on a local criterion (more details about the criterion are discussed in Section~\ref{sec:STHMC}).

\subsection{Molecular Dynamics}
While in MC of a classical system, one is only concerned with the (potential) energy of the system given by the particle positions $\bm{x}$, in MD one simulates the combined phase space of coordinates and velocities $\bm{v}$, i.e., one has $\bm{\xi}=(\bm{x},\bm{v})$.
Further, in standard MD, one simulates in the microcanonical ensemble, manifesting in principally conserved total energy $H(\bm{\xi})=H(\bm{x},\bm{v})=U(\bm{x})+K(\bm{v})$, where the potential energy $U(\bm{x})$ depends only on the coordinates $\bm{x}$ and the kinetic energy $K(\bm{v})$ depending on the velocities $\bm{v}$.
This is achieved by iteratively integrating the equation of motions, where the most common prescription used is the velocity Verlet algorithm,\cite{swope1982computer} consisting of the following steps
\begin{subequations}
\begin{eqnarray}
    \bm{x}_i(t_i+\Delta t_i) = \bm{x}_i(t_i)+\bm{v}_i\left(t_i\right)\Delta t_i+\frac12 \bm{a}_i(t_i) \Delta t_i^2\\
    \bm{v}_i\left(t_i+\Delta t_i\right) = \bm{v}_i(t_i) + \frac{\bm{a}_i(t_i) +\bm{a}_i(t_i+\Delta t_i)}{2} \Delta t_i
\end{eqnarray}
\label{eq:vel_ver}
\end{subequations}
\noindent
where $i$ is the index of the atom, $\Delta t_i$ is the (atom dependent) timestep, and $\bm{a}_i=-m_i^{-1}\partial/\partial \bm{x}_i U(\bm{x})$ is the acceleration acting on the atom obtained from the potential.
In a standard MD simulation, the individual atoms need to evolve synchronously in time, which practically restricts the use of timestep to a global definition of $\Delta t_i=\Delta t$, i.e., the timestep is not dependent on the atom index.
That also implies that the maximal timestep which can be used is determined by the fastest mode of oscillation.
As we discuss later, this restriction is not necessary for HMC, being one source of potential speed-up compared to MD.
\par
To sample from $P^{\mathrm{eq}}$ in the canonical ensemble using MD, one has to additionally find a method to control the velocities, for which there is no natural way.\cite{hunenberger2005thermostat}
Each of the established approaches has certain advantages and disadvantages.
In the canonical ensemble for MD, one commonly reproduces the Boltzmann distribution of the total energy
\begin{equation}
    P^{\mathrm{eq}}(\bm{\xi}) = P^{\mathrm{eq}}(\bm{x},\bm{v})=\frac{e^{-H(\bm{x},\bm{v})/k_BT}}{\int e^{-U(\bm{x})/k_BT} d\bm{x} \int e^{-K(\bm{v})/k_BT} d\bm{v}},
    \label{eq:boltz_total}
\end{equation}
which factorizes into the canonical distribution of the potential energy and of the momenta
\begin{equation}
    \begin{split}
    P^{\mathrm{eq}}(\bm{x},\bm{v})&=\frac{e^{-U(\bm{x})/k_BT}}{\int e^{-U(\bm{x})/k_BT} d\bm{x}} \frac{e^{-K(\bm{v})/k_BT}}{\int e^{-K(\bm{v})/k_BT} d\bm{v}} \\
    &=P^{\mathrm{eq}}(\bm{x})P^{\mathrm{eq}}(\bm{v}).
    \end{split}
    \label{eq:boltz_total_fact}
\end{equation}
Not all thermostats produce the canonical ensemble, so special care has to be taken to make the right choice.\cite{hunenberger2005thermostat}
Especially, some thermostats are only canonical in $P^{\mathrm{eq}}(\bm{x})$, but not in the joint distribution $P^{\mathrm{eq}}(\bm{x},\bm{v})$.
\par
We also want to highlight that sampling using MD is only approximate, i.e., the convergence to the target distribution is only guaranteed in the limit of $\Delta t_i \rightarrow 0$.
In contrast, MC sampling is asymptotically exact.

\subsection{Hamiltonian Monte Carlo}
\label{subsec:HMC}
HMC combines elements from MD and MC: Microcanonical MD simulations are used as proposals for the MC accept/reject step.
This combination of methods was originally proposed by Duane et al.\cite{duane1987hybrid} and later popularized in the statistics community with applications towards inference of Bayesian neural networks.\cite{neal2011mcmc}
There exists recent work highlighting the performance of HMC,\cite{prokhorenko2018large} and it is implemented in or for commonly used simulation packages.\cite{hu2006monte,fernandez2014constant,john_chodera_2022_6958059}
However, one major hurdle is choosing the optimal parameters of the simulation, as will become clear in the following.
\par
For this method, we again have both particle positions and velocities as our state, i.e., $\bm{\xi}=(\bm{x},\bm{v})$.
The steps of HMC are as follows:
\begin{enumerate}
    \item Draw velocities $\bm{v}_k$ according to Maxwell-Boltzmann distribution, generating the initial state $\bm{\xi_k}=(\bm{x}_k,\bm{v}_k)$. This sets a new level of total energy $H(\bm{\xi}_k)$.
    \item Propagate the system according to Eqs.~(\ref{eq:vel_ver}) for $n$ steps with fixed $\Delta t_i=\Delta t$, resulting in a proposal configuration $\bm{\xi}_l$. The integration steps are performed in the microcanonical ensemble, corresponding to principally conserved total energy $H$ (in practice, this is not the case due to the discretization in time).
    \item The new state $\bm{\xi_l}$ is then accepted according to Eq.~(\ref{eq:metropolis}). If the proposal is rejected, the system is reset to $\bm{\xi}=\bm{\xi}_k$ and one continues at step 1.
\end{enumerate}
The overall prescription produces a canonical distribution of the total energy, which following Eq.~(\ref{eq:metropolis}) (with $f(\xi_k,\xi_l)=f(\xi_l,\xi_k)$) produces our target distribution of Eq.~(\ref{eq:peq}).
There are some important details of this procedure which we will discuss in the following.
\par
It has been realized in the HMC literature that it is beneficial to jitter $\Delta t_i$, i.e., to not fix $\Delta t_i$ but to pick it from some distribution, to avoid some problems related to repeatedly running into small unfavorable regions in phase-space due to the deterministic dynamics.\cite{neal2011mcmc}
More details on this will be discussed in Section~\ref{sec:STHMC}.
\par
HMC replaces (or actually can augment) the hand-crafted move sets used in a standard MC simulation.
The advantage of this approach is the additional use of the forces to propose the moves, which allows for informed moves that either dissipate or absorb kinetic energy.
This way, the acceptance probability is drastically improved while maintaining relatively large conformational changes.

\subsubsection{Choice of Timestep and Number of Integration Steps}
For an effective exploration of phase space, one needs to balance phase-space movement and acceptance rates.
Indeed, it has been shown in Ref.~\onlinecite{beskos2013optimal} that the optimal acceptance probability should approach $65.1\%$ for HMC (under some assumptions using the standard leapfrog integrator), independent of the particular (high dimensional) target.
In practice, however, it is not clear if all assumptions hold and it has been found to sometimes perform poorly, especially due to the observation that samplers with the same acceptance rate can exhibit vastly different behavior.\cite{wang2013adaptive}
\par
The often employed practical solution of tuning the parameters of HMC is thus to prescribe a target acceptance rate of around $60\%$ and heuristically optimize the timestep $\Delta t$ to take such a value that this is observed on average.
A common method to set a good number of integration steps $n$ is called NUTS,\cite{hoffman2014no} which uses a recursive algorithm to build a set of candidate configurations on the fly.
This procedure stops once the so-called U-Turn condition is satisfied, which signifies a doubling back of the trajectory.
While this approach works well in practice, to preserve detailed balance, there is a need to both consider moves in forward and backward directions which recursively build a tree of steps, from which one then samples the proposal.
This can incur a two-fold overhead in performed updates, resulting in wasted computation time.
\par
Reference~\onlinecite{hoffman2021adaptive} proposed an improved gradient-based approach to tune HMC: The timestep is tuned based on a target acceptance probability, while the number of integration steps was optimized based on a local objective incorporating multiple Markov chains.
This way, the approach can tune the total trajectory length $n\Delta t$, i.e., to some extent to tune both crucial parameters of HMC.
There, it is shown that such an approach can outperform NUTS and find the optimal parameters one would otherwise find through an extensive and costly grid search.

\subsubsection{Choice of Integrator}
The propagation of the system can be performed by any arbitrary function and does not need to follow the structure set in Eqs.~(\ref{eq:vel_ver}) for the velocity Verlet algorithm.
For example, one could use higher-order integrators, such as variants of the Runge-Kutta algorithm.\cite{butcher2016numerical}
In particular, for the velocity Verlet algorithm, as already hinted at, $\Delta t_i$ can be atom-dependent (or even be independent for each degree-of-freedom), not simulating within the microcanonical ensemble any longer.
This may seem counterintuitive, but from the point of view of generating a trial configuration for an MC simulation, there is no requirement for the integrator to reproduce microcanonical trajectories.
Indeed, as we will show in Section~\ref{sec:atom_dependent}, this simple extension can lead to an acceleration of sampling.
\par
A similar approach was originally introduced in Ref.~\onlinecite{l2hmc} by using neural networks to reparameterize the integrator.
There, the parameters of the neural network were optimized using a gradient-based optimization approach using a local loss (see next section) for a set of statistical distributions and latent-variable generative models.
However, in their set-up, they are not able to learn $n$ and did not optimize $\Delta t$.
\par
For general integrators as for example used in Ref.~\onlinecite{l2hmc}, the change in phase-volume needs to be accounted for in the accept/reject step of Eq.~(\ref{eq:metropolis}).
We refer to the literature on normalizing flows for details on this, in particular Refs.~\onlinecite{tabak2010density,tabak2013family,papamakarios2021normalizing}.
For the case of atom-dependent timesteps, the change in phase-volume is zero, so that we can simply use the ratio of our target distribution $P^{\mathrm{eq}}(\bm{\xi})$ and do not need to account for this explicitly.

\section{Self-tuning Hamiltonian Monte Carlo}
\label{sec:STHMC}
The goal of our work is to present a generally applicable approach that allows for a gradient-based optimization of simulation parameters of HMC for molecular simulations.
In particular, we will focus on optimizing the timesteps $\Delta t_i$ (both global and atom-based) and the number of steps $n$ used in HMC.
Such an approach has three major benefits: \emph{i}) It eliminates the otherwise needed expensive grid search over these parameters to arrive at good simulation parameters, \emph{ii}) it can additionally speed up the simulation when using atom-based timesteps, and \emph{iii}) it is easy to implement in machine-learning frameworks with automatic differentiation, such as PyTorch.\cite{NEURIPS2019_9015}
\par
In the following, we present the general simulation setup and discuss how gradient-based optimizers can be utilized to find good parameters of our simulation based on a local definition of the loss, as proxy for the autocorrelation time, that promotes phase-space exploration.
We also discuss the importance of avoiding local minima for the optimization by not considering timesteps that are fixed, but picked from a distribution.
Compared to the approaches presented in the last section, our approach allows the combined learning of $\Delta t_i$ and $n$ without any additional assumptions.

\subsection{Fully Differentiable Simulation Set-Up}
\label{sec:fully_diff}
Gradient-based optimizers are known to be very efficient in finding minima in high dimensional problems, making them a prime candidate for our approach since the number of atoms can grow quite large, constituting many correlated parameters.
Although classical gradient-based optimizers, by definition, only use local information and can be trapped in a local minimum, we show that in our application they can very efficiently find suitable values for the parameters.
\par
We achieve a fully differentiable simulation set-up by implementing our algorithm in PyTorch,\cite{NEURIPS2019_9015} a software library often used in machine learning.
The appeal of this approach is automatic differentiation\cite{rall1981automatic,paszke2017automatic} which allows for the evaluation of partial derivatives of a function specified by a computer program.
In PyTorch, any operation applied to the so-called tensors is recorded, so that via the chain rule one can calculate the gradient on any parameter of the computation graph.
To make changes based on this gradient information, we then need a way to judge the goodness of the output of the computation, i.e., a loss $L$ (details discussed in the next section) associated with the integration.
\par
The parameters are then updated depending on the loss using backpropagation,\cite{rosenblatt1961principles,rumelhart1985learning} i.e., the derivative of the loss with respect to every parameter $\theta_m$ of the computation graph we want to tune is calculated and then used to update the value of the parameter
\begin{equation}
    \theta_m' = \theta_m -\eta \frac{\partial L}{\partial \theta_m},
\end{equation}
where $\eta$ is the so-called learning rate and $\theta_m$ is for example the timesteps $\Delta t_i$ or the number of integrations steps $n$.
This type of update rule is also called gradient descent, i.e., the parameters are changed in the opposite direction of the gradient.
Variants of this optimization algorithm exist, which for example also include a momentum variable to accelerate convergence.
We will use one of the most popular optimizers incorporating such additional terms, i.e., the Adam optimizer.\cite{kingma2014adam}
\par
\subsection{Autocorrelation and Loss Definition}
In the following, we discuss the definition of the autocorrelation function/time and how we propose to define a proxy loss for it.

\subsubsection{Autocorrelation}
\label{sec:obj}
The most common way to evaluate the performance of an MC simulation is to investigate the autocorrelation between subsequent states of the chain.
For this, we first recall the definition of the autocorrelation function\cite{janke2013monte} 
\begin{equation}
    A_{\mathcal{O}}(k)=\frac{\left\langle\mathcal{O}_i \mathcal{O}_{i+k}\right\rangle-\left\langle\mathcal{O}_i\right\rangle\left\langle\mathcal{O}_i\right\rangle}{\left\langle\mathcal{O}_i^2\right\rangle-\left\langle\mathcal{O}_i\right\rangle\left\langle\mathcal{O}_i\right\rangle},
\end{equation}
where $k$ is the lag time, $\mathcal{O}$ is any observable of the system, and $\langle \ldots \rangle$ symbolizes the thermodynamic expectation value in equilibrium when sampling $P^{\mathrm{eq}}$.
From the autocorrelation function, one way to obtain the autocorrelation time is
\begin{equation}
    \tau_{\mathcal{O}}=\frac{1}{2}+\sum_{k=1}^{N_t} A_{\mathcal{O}}(k)\left(1-\frac{k}{N}\right),
    \label{eq:sum_auto}
\end{equation}
where $N_t$ is the number of measurements.
The autocorrelation time is related to the effective sample size (ESS) 
\begin{equation}
N_{\mathcal{O},\mathrm{eff}}=\frac{N}{2\tau_{\mathcal{O}}}.
\end{equation}
In practice, we calculate the ESS as implemented in tensorflow\cite{dillon2017tensorflow} where the sum in Eq.~(\ref{eq:sum_auto}) is truncated as proposed in Ref.~\onlinecite{geyer1992practical}.
We then use the ESS to estimate the autocorrelation time $\tau_{\mathcal{O}}$ via the above relation.
\par
The importance of the autocorrelation time lies in the need to be included when calculating the standard deviation on observables as
\begin{equation}
    \sigma_{\overline{\mathcal{O}}}^2=\frac{\sigma_{\mathcal{O}}^2}{N_{\mathcal{O},\mathrm{eff}}}=\frac{\sigma_{\mathcal{O}}^2}{N} 2 \tau_{\mathcal{O}}.    
\end{equation}
That is, when an algorithm has a smaller autocorrelation time one needs to simulate shorter to achieve the same error on the observable.

\subsubsection{Proxy Loss}
Quantities related to the autocorrelation function cannot be effectively used as an objective for the fully-differentiable set-up since they require long chains to provide reliable estimates of $\tau_{\mathcal{O}}$.
Therefore, we make use of a proxy loss to the autocorrelation time, defined as
\begin{equation}
    L_n = - p_{n} |\bm{x'}_n-\bm{x}_0|^b
    \label{eq:loss}
\end{equation}
where $p_n$ is the acceptance probability of the proposal and $|\bm{x'}_n-\bm{x}_0|^b$ is the movement in coordinate phase-space (distance between the start and end states), computed after performing $n$ integration steps, i.e., for each proposal we generate.
A common choice in the adaptive MC literature is to use $b=2$,\cite{beskos2009optimal,pasarica2010adaptively,bojesen2018policy} where one is thus optimizing for the expected squared jump distance.
This definition, however, is not unique and is not guaranteed to provide the best correspondence to a reduction in autocorrelation times for all observables.
For example, an optimal jump in real coordinates does not need to lead to an optimal autocorrelation time for other observables such as the potential energy.
Further, the expected jump distance only optimizes for the lag-1 autocorrelation, whereas further values are ignored.
It is not clear how well the information about correlations at small lags correlates with the overall shape of the autocorrelation function.
This means that there is still some freedom in optimizing the loss function.
\par
For example, in Ref.~\onlinecite{l2hmc} the authors introduced an additional reciprocal term that penalized small jumps more strongly, and in Refs.~\onlinecite{hoffman2021adaptive,sountsov2021focusing} alternative definition relying on multiple chains are proposed, either by evaluating the change in the estimators of the expected squared jump\cite{hoffman2021adaptive} or focusing on difficult directions.\cite{sountsov2021focusing}
As we will show later, for alanine dipeptide we found that simply setting $b\approx 4 >2$ provides a better correspondence with the autocorrelation times we observe for the potential energy.
\par
Since our goal is to propagate the system as efficiently as possible in terms of computational effort, the definition of the loss in Eq.~(\ref{eq:loss}) is only sufficient when the number of integration steps $n$ is fixed.
For our purposes, we introduce a rescaling of the loss by the computational effort, i.e., we define the loss as $L_n/n$ and by this incorporate the information that every integration step takes roughly the same computational effort.
Previously, it was empirically found that defining $L/\sqrt{n}$ provides a well-working approach in practice,\cite{wang2013adaptive} although it is not entirely clear to us why the cost should not enter linearly.
\subsubsection{Learning the Number of Integration Steps}
When optimizing for the number of integration steps one needs to find a way to include the information about them in the loss for each $n$, which allows calculating partial derivatives with respect to it and gives a signal for good values.
Thus, to learn the optimal (distribution of the) number of integration steps $n$, we propose to weight the output of every integration step with a learned distribution.
For this, we define the loss as 
\begin{equation}
    L = \sum_{n=1}^N c_n L_n / n,
\end{equation}
where $c_n$ are the weights of the particular number of integration steps and $N$ is the maximal number of integration steps considered during training.
The $c_n$ are in practice obtained as softmax of unrestricted parameters $C_n$, i.e.,
\begin{equation}
    c_n = \sigma(C_n) = \frac{e^{C_{n}}}{\sum_{n=1}^N e^{C_{n}}},
\end{equation}
where the temperature of the softmax is set to unity.
We initialize $C_n$ as uniform random numbers from zero to one and apply the softmax to arrive at our initial $c_n$.
\par
This approach is inspired by an attention-like set-up\cite{vaswani2017attention} and allows to give ``attention'' towards a particular integration step.
While this approach makes it necessary to set a maximum number of integration steps $N$ during training and always simulate until the maximum is reached, after learning one categorically picks from the probabilities and there is thus no unnecessary computing.
\par
In practice, we also normalize the loss by the number of atoms (usually fixed during a simulation) and the number of epochs to arrive at optimal learning rates that are as independent of the system as possible.
Since they are only constant factors during training, these do, however, do not influence the system apart from rescaling the learning rate.

\subsection{Local Minima and Jittering}
\label{sec:jittering_local}
As noted before, local minima in the loss are a problem for classical gradient-based optimizers.
This is a potential limitation of our approach when optimizing parameters of HMC, as their existence cannot always be ruled out.
\par
One known source of problems in HMC is using a fixed timestep which can potentially lead to problems related to deterministically sampling unfavorable configurations.
Jittering of the timesteps $\Delta t_i$ is a well-known approach to avoid this potentially detrimental periodic behavior of the integrator.
In our case, we pick $\Delta t_i$ from a normal distribution with fixed relative variance, i.e., 
\begin{equation}
    \Delta t_i' \sim \mathcal{N}(\Delta t_i,s\Delta t_i),
    \label{eq:jitter_dt}
\end{equation}
where $s$ is a free parameter.
As we will show in Section~\ref{sec:harmonic} for the harmonic oscillator, the introduction of jittering has particular importance when optimizing based on the local proxy loss using gradient-based optimizers, which has to our knowledge not been realized before.
Without jittering several local minima occur in the loss landscape of $\Delta t$ and $n$ trapping the optimization there.
We thus show that jittering avoids one source of local minima in the loss.
For high dimensional optimization problems, this may not be the only source of local minima in the loss surface.
In such a setting, it is possible to probe whether local minima exist by starting with different initial parameter guesses and checking whether the parameters converge to the same values after optimization, but not to investigate the loss surface systematically.
While we cannot solve the global optimization problem, our setup allows at least the optimization of parameters within the basin of attraction.
This is indeed what we observe when simulating alanine dipeptide with atom dependent timesteps $\Delta t_i$ in Section~\ref{sec:atom_dependent}, for which we find that the optimization for some initial parameters appears to get trapped in a (worse) local minima.

\section{Results}
\label{sec:res}
We first study the one-dimensional Harmonic oscillator in our framework.
This allows us to highlight some important aspects, such as the occurrence of shadow Hamiltonians for large timesteps and the periodic behavior in the loss, leading to multiple minima when not using jittering of the timestep.
Once we have understood the peculiarities of our approach, we explore the use of our self-tuning HMC framework on the physically more realistic protein system, alanine dipeptide.
Here, we focus on the definition of the loss, in particular, the ideal value of $b$ in Eq.~\ref{eq:loss} for this class of systems.
We extend the general approach that allows the replacement of the grid-search by including atom-dependent timesteps, which in our case leads to a further speed-up without additional overhead.

\subsection{Harmonic Oscillator}
\label{sec:harmonic}
\begin{figure}
    \includegraphics{"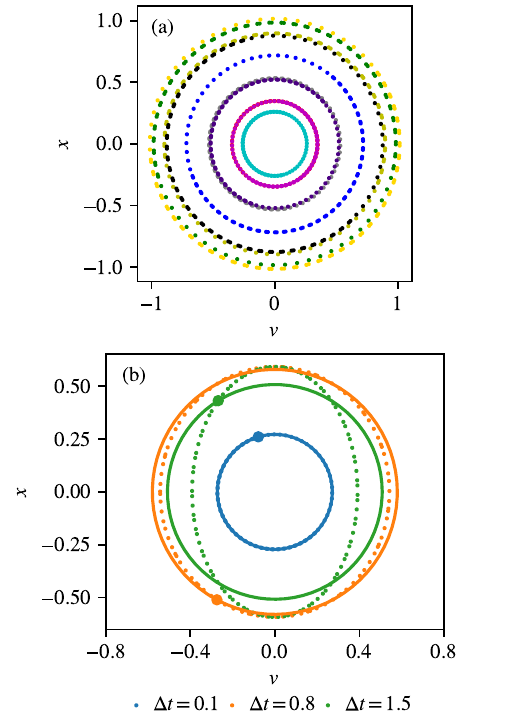"}
    \caption{(a) Example trajectories for $x$ and $v$ of the one-dimensional harmonic oscillator for $\Delta t=0.1$ at $T=0.5$ as obtained from HMC with $n=100$. (b) Influence of the choice of $\Delta t$ on the simulated (shadow) Hamiltonian using otherwise the same parameters as in (a). The solid lines in the same color as the data points correspond to the analytically expected trajectories. The big dots symbolize the starting point of the trajectory, which sets the expected energy level.}
    \label{fig:phase_space_harmonic}
\end{figure}

The full Hamiltonian of the one-dimensional harmonic oscillator with mass $m=1$ and spring constant $k=1$ is defined as
\begin{equation}
\mathcal{H} = U(x) + K(v) = 0.5(x^2+v^2),
\label{eq:Ham_Harmonic}
\end{equation}
where $x$ is the position and $v$ is the velocity of the mass.
Our goal for this exemplary system is to simulate it at a fixed temperature in the canonical ensemble, for which we choose $T=0.5$ here ($k_B=1$ in this case).
While this is one of the most simple systems one can consider and has been considered as a test system in some cases,\cite{neal2011mcmc} a systematic study in the framework of HMC is lacking, especially in the context of a self-tuning approach.
\subsubsection{Phase Space and Simulated Shadow Hamiltonian}
In Fig.~\ref{fig:phase_space_harmonic}(a) we visualize a few sample trajectories of the harmonic oscillator for $\Delta t=0.1$ and $n=100$ integration steps obtained from HMC.
The trajectories form (near perfect) circles, where the radius is given by energy conservation of the Hamiltonian~(\ref{eq:Ham_Harmonic}).
The different radii of the circles can be understood since at every new iteration the velocity is picked from the Maxwell-Boltzmann distribution, setting a different level of the total energy.
With these parameters, the total energy is nearly perfectly conserved for each trajectory, leading to acceptance rates close to $100\%$.
\par
It is well known in the literature that, when using a finite timestep, one only simulates the so-called shadow Hamiltonian\cite{engle2005monitoring,neal2011mcmc,zolotov2013accurate, kim2015time} and not the true Hamiltonian.
The difference between these two Hamiltonians is dependent on the chosen timestep $\Delta t$ and can readily be observed for the harmonic oscillator, for which we plot example trajectories for different $\Delta t$ in Fig.~\ref{fig:phase_space_harmonic}(b).
While for $\Delta t=0.1$, the circles are perfect on this scale as before, for $\Delta t=0.8$ and $1.5$~fs the trajectory clearly forms an ellipse with eccentricity $e>0$.
As solid lines in the same color as the data points, we have also drawn the trajectories that should theoretically have been simulated, simply following the energy conservation prescribed by the Hamiltonian (\ref{eq:Ham_Harmonic}).
These true trajectories start from the initial point of the trajectory marked by the big dot in the same color.
It can be seen that the points deviate more from the true circle for increasing $\Delta t$.
Whenever one observes points on the inside of the circle, this corresponds to lower total energy and thus acceptance of $100\%$, whereas points outside the circle have larger total energy and are not always accepted.
\par
While one would naively expect that these effects should average out since one often starts from a new initial position, resulting in an average acceptance rate during a full HMC simulation, this is not the case.
Since the first point of the phase-space sets the initial energy level (which in turn sets the radius of the circle) and one always moves along the trajectory the same distance (given by $n\Delta t$) in either forward or backward direction, one sees a periodic behavior of the acceptance rate given by the ``deviation from the circle'' in Fig.~\ref{fig:phase_space_harmonic}.
This has crucial effects on our approach, as shown in the next sections.
\subsubsection{Influence of Jittering}
\begin{figure}
    \includegraphics{"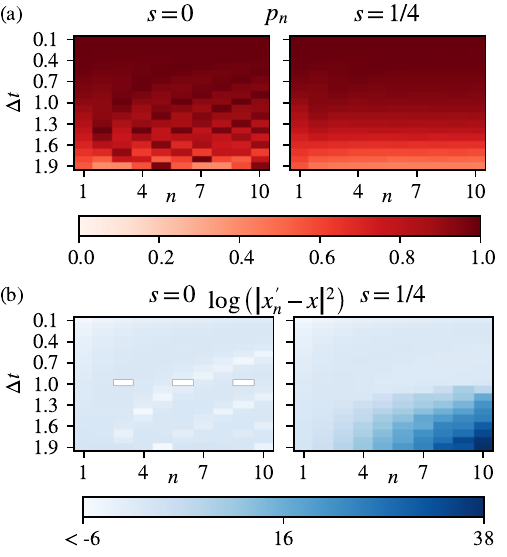"}
    \caption{(a) Acceptance $p$ and (b) logarithm of squared jump $(x'_n-x)^2$ as a function of timestep $\Delta t$ and number of integration steps $n$ for the one-dimensional harmonic oscillator at $T=0.5$.
    Shown are in both cases the results for the not jittered ($s=0$) and jittered ($s=1/4$) timestep $\Delta t$.}
    \label{fig:jitter_vs_no_jitter}
\end{figure}
In this section, we will investigate the advantage of jittering the timestep $\Delta t$ on the dynamics of the harmonic oscillator as an approach to avoid recurring patterns, as observed in the last section.
We jitter the timestep following the definition in Eq.~(\ref{eq:jitter_dt}).
Figure~\ref{fig:jitter_vs_no_jitter}(a) shows a heatmap of the acceptance rate $p_n$ as a function of timestep $\Delta t$ and the number of integration steps $n$ for the harmonic oscillator, both without ($s=0$) and with jittering ($s=1/4$), measured after the system is equilibrated.
Without jittering one observes several minima/maxima in the surface plot, corresponding to small/large acceptance rates.
They follow a pattern, which can exactly be explained by the deviations from the true trajectories discussed in Fig.~\ref{fig:phase_space_harmonic}(b).
\par
When introducing jittering on the timestep, as shown in the same plot where we have used $s=1/4$, these minima/maxima vanish and one observes as a function of $\Delta t$ a smooth decay of the acceptance rate $p_n$.
This decay is also there in the non-jittered simulation, but less visible due to the overlay with the many minima/maxima.
\par
This highlights the problem of using a simple criterion of fixed target acceptance rate, as often done when tuning the parameters of HMC.
Without jittering, one would pick, depending on the starting parameters, any pair of $\Delta t$ and $n$ having the desired value of acceptance rate, which does not need to correspond to a small autocorrelation time (as shown in the next section).
For the jittered simulations, one would pick a fixed $\Delta t$ as local minima/maxima are smoothed out, but without any ability to distinguish between the influence of $n$ on the performance.
As we will see later, this does not correlate well with the autocorrelation times of the potential energy.
\par
A similar behavior of multiple local minima/maxima can also be observed for the squared jump distance, presented in Fig.~\ref{fig:jitter_vs_no_jitter}(b).
Here, we have opted to plot it logarithmically, since the differences in the jump are quite large for some parameter configurations.
Without jittering ($s=0$), one observes several minima/maxima in the surface plot, whereas with jittering $s=1/4$ this is not seen.
With jittering, however, finds that the squared jump distance can become huge for the larger $\Delta t$ and $n$ region, which can be explained by the occasional ``breaking'' of simulations at large $\Delta t$ where self-enforcing effects lead to explosions of the values of the position and velocity.
This is a well-known effect when choosing very large time steps and are typically rejected by the Metropolis-Hastings criterion due to a very large potential energy.
\subsubsection{Loss Surface}
\begin{figure*}
    \includegraphics{"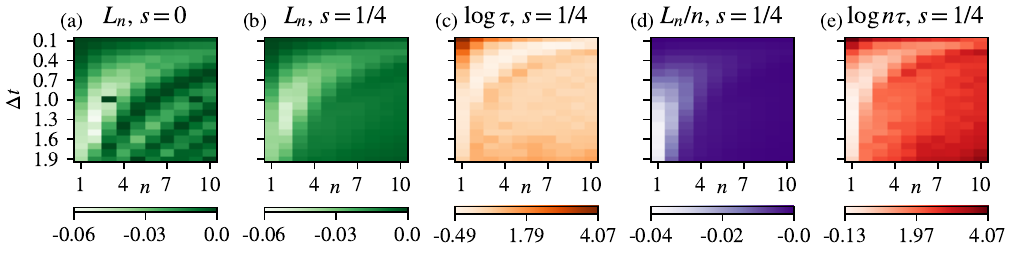"}
    \caption{(a) and (b) show the loss $L_N$ as a function of $\Delta t$ and $n$ for (a) no jittering ($s=0$) and (b) with jittering ($s=1/4$). In (c), we plot the logarithm of the autocorrelation time extracted from the time-series of the potential energy.
    The region of desired small aucorrelation times corresponds reasonable well to the region where the loss is minimized, as shown in (b).
    (d) shows the same data as in (b), but the loss is rescaled with the computational effort $L_n/n$. Finally, in (e) we again show the logarithm of the autocorrelation time, but in units of the computational effort $n\tau$.
    The logarithm is chosen for the autocorrelation time to highlight the differences, as these are much larger than in the other plots for the losses.}
    \label{fig:harmonic_loss}
\end{figure*}
The observations from the last section have crucial effects on our definition of the loss.
Figure~\ref{fig:harmonic_loss}(a) and (b) show the loss $L_n$ of Eq.~(\ref{eq:loss}) recorded during the HMC run (\emph{without learning}, i.e., only showing the obtained values for the optimization target).
The general observation of multiple minima/maxima for the separate acceptance rate and squared jump distance in Fig.~\ref{fig:jitter_vs_no_jitter}(a) and (b) also carries over to the loss (correlated expectation value of both) without jitter, see Fig.~\ref{fig:harmonic_loss}(a).
With jitter, as shown in Fig.~\ref{fig:harmonic_loss}(b), the loss loses this property, and one (clear) global minimum emerges at $\Delta t\approx 1.3$ and $n\approx2$.
\par
Our goal is that the loss serves as a local proxy for the autocorrelation times of our observables, where we here focus on the correlations of the potential energy as a placeholder for many interesting properties of the system.
While not optimally, the loss agrees generally well with the (logarithms of the) autocorrelation times for the potential energy presented in Fig.~\ref{fig:harmonic_loss}(c).
The region with lower loss appears to be shifted relatively towards higher $\Delta t$, which however is not as detrimental as smaller $\Delta t$.
Finally, we are, however, interested in the performance per computing effort.
For this, we plot $L_n/n$ in Fig.~\ref{fig:harmonic_loss}(d), which shifts the minimum towards smaller $n$.
The global optimum for this system is somewhere around $\Delta t\approx 1.75$ and $n\approx 1$.
This is also reconfirmed for the autocorrelation time $n\tau$ shown in Fig.~\ref{fig:harmonic_loss}(h), measured in terms of the computational effort.
Here, we also observe that the minima shift towards smaller $n$, although not as strongly as for the loss.
This is due to the relative difference in the amplitude between minimum/maximum for the loss and autocorrelation time.
\par
The difference in autocorrelation times even for this simple system is quite large, corresponding roughly to a difference of $100$, highlighting the importance of choosing suitable parameters for $\Delta t$ and $n$. 

\subsubsection{Learning of HMC Parameters}
\begin{figure*}
    \includegraphics{"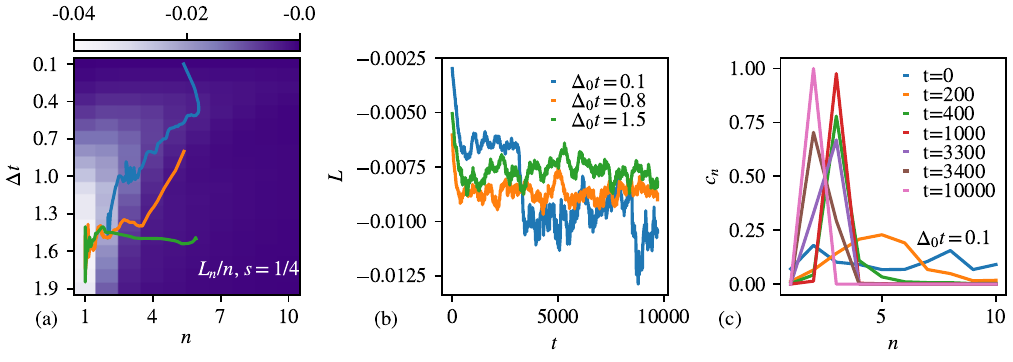"}
    \caption{(a) Loss surface as a function of $\Delta t$ and $n$. On top, three example trajectories show the expectations values of $\Delta t$ and $n$ during learning for three initial values of the timestep $\Delta_0 t$. (b) The corresponding loss as a function of epochs $t$ for the curves shown in (a). (c) Attention weights $c_n$ for $\Delta_0 t=0.1$ for different epochs indicated in the legend.}
    \label{fig:harmonic_learning}
\end{figure*}
We have now established a good correlation between the loss and the autocorrelation time of the potential energy for this system and shown that local minima in the loss surface are eliminated by jittering the timestep.
With this setup in place, we now turn to learn the optimal parameters via the fully differentiable framework, implemented in pyTorch.\cite{NEURIPS2019_9015}
We make use of the Adam optimizer\cite{kingma2014adam} with learning rate $\epsilon=0.01$ with otherwise default parameters from PyTorch.
Before performing an optimization step of $\Delta t$ and $n$, we perform $10$ proposals to average out the resulting gradients, setting our epoch length.
The system is initialized for different $\Delta_0 t$ and our attention weights $C_n$ are picked randomly uniform from zero to one (this means our mean value of the number of steps $n$ is initially $\approx N/2$ and allows for ``information'' from all integration steps).
\par
In Fig.~\ref{fig:harmonic_learning}(a) we again show the loss surface of $L_n/n$ presented in Fig.~\ref{fig:harmonic_loss}(d), but now include sample learning trajectories for different initial values of $\Delta_0 t$ with the $C_n$ randomly initialized as discussed before.
We plot the mean values of the timestep and number of integration steps, i.e., $\Delta t$ and $\overline{n}=\sum_{i=1}^{N}nc_n$.
It is evident from the visualization of the trajectories that, independent of the initial parameter, all curves move towards the region with smaller loss values.
This is reinforced by the recorded loss values during training, which we plot as a function of training epoch $t$ in Fig.~\ref{fig:harmonic_learning}(b).
Since the data for the recorded loss is very noisy due to the small batch size and few degrees of freedom of the system leading to little self-averaging, we have calculated a running average over $300$ epochs to visualize the training.
\par
For $\Delta_0 t=0.1$ some jumps in the loss are visible, mainly around $t=3000$.
This corresponds to the ``jump'' from a big weight at $n=3$ to $n=2$, which can also be appreciated from Fig.~\ref{fig:harmonic_learning}(c), where we show the attention weights $c_n$ for different epochs.
Initially, at $t=0$, the weights are nearly uniform, whereas then for early times at $t=200$ a small peak forms around $n=5$.
As time progresses, there is a large weight on $n=3$ which then in the time-span from $t\approx1000$ shifts towards $n=2$ at $t\approx 3400$.
At the final training epoch, the weight is completely on $n=2$.
These observations are consistent with what we see in Fig.~\ref{fig:harmonic_learning}(a) as movements on the loss surface for $\Delta_0 t=0.1$.
\par
Having a good correspondence between loss and the autocorrelation of the potential energy we have shown that our fully differential framework allows us to effectively learn good parameters of HMC without the need for an expensive grid search.
Next, we will consider a bigger molecular system with more intricate interactions.

\subsection{Alanine Dipeptide}
\begin{figure}
    \centering
    \includegraphics[width=0.4\textwidth]{"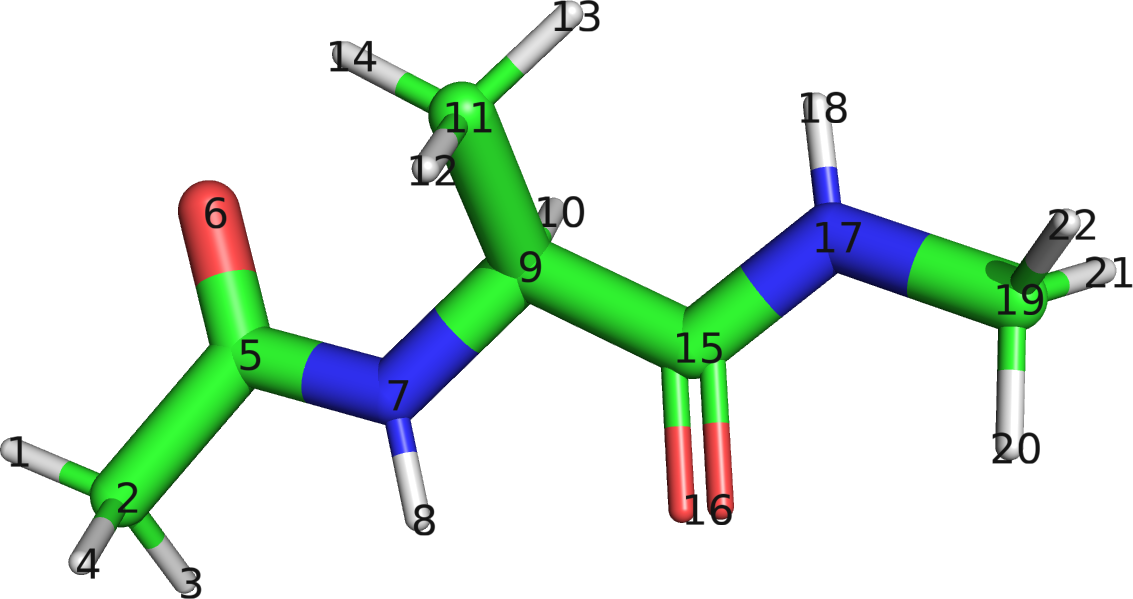"}
    \caption{Graphical representation of alanine dipeptide, where the atoms are marked by their index. The white color symbolizes hydrogens, the green color stands for carbon, the red color represents oxygen and blue is for nitrogen.}
    \label{fig:graph_ala}
\end{figure}

Alanine dipeptide has proven itself as the most common protein to test novel algorithms, which is why we investigate it in the following.
We simulate this system in vacuum, thus this protein has $66$ positional degrees-of-freedom ($22$ atoms in $d=3$ spatial dimensions) with some interaction between atoms being bonded, as drawn schematically in Fig.~\ref{fig:graph_ala}.
As force-field, we make use of Amber-19ffSB, for which we have adapted the implementation from TorchMD\cite{doerr2021torchmd} for our purposes.\footnote{In the original implementation, the gradients calculated by automatic differentiation needed to make this problem fully-differentiable are not preserved. We have adapted the code to preserve those.}
We note that to use gradient-based optimizers, we have to differentiate through the whole computation graph, including the integrator and the force field.
For a description of the functional form of the potential energy part of the Hamiltonian, we refer to Ref.~\onlinecite{tian2019ff19sb}.
The temperature is set to $T=300$~K and free boundary conditions are employed.
For the HMC simulations, we have found that jittering with $25\%$ can lead to non-stable simulations for larger $\Delta t$, which is why we here chose to use a jitter with $10\%$ relative variance, i.e., $s=0.1$.
The loss and the autocorrelation time surfaces were calculated with fixed parameters after equilibration and obtained from time-series with $2 \times 10^5$ MC proposals.

\subsubsection{Adaptation of the Loss}
\begin{figure*}
    \includegraphics{"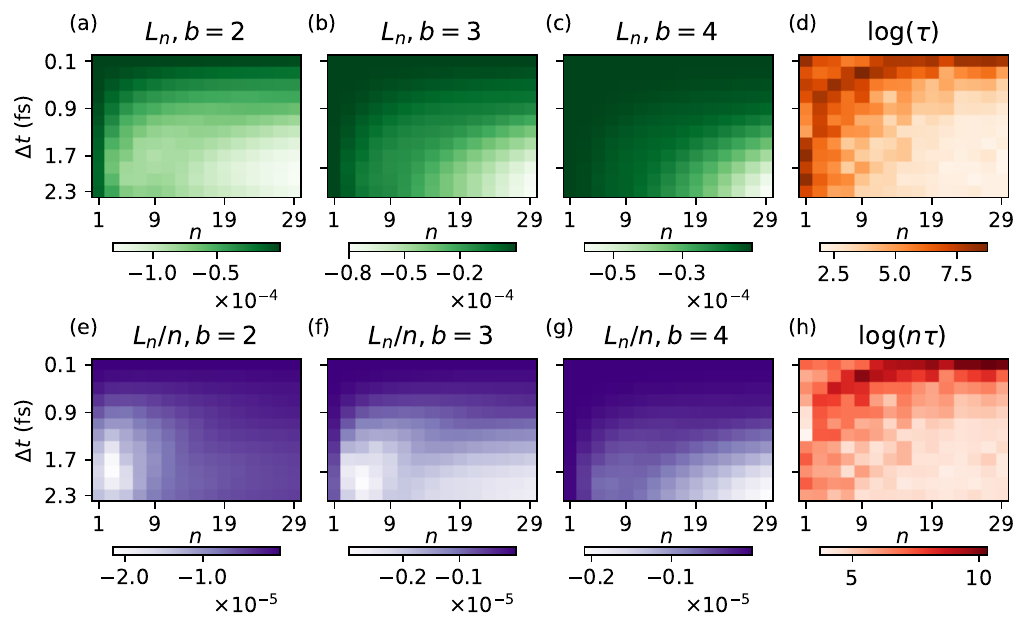"}
    \caption{Influence of the parameter $b$ in the loss of Eq.~(\ref{eq:loss}) on the loss is shown for alanine dipeptide for (a) $b=2$, (b) $b=3$, (c) $b=4$ on the loss surface $L_n$ as a function of $\Delta t$ and $n$ is shown.
    In (e) the corresponding logarithm of the autocorrelation time of the potential energy is presented.
    (e)-(g) show the corresponding plots of the loss per computational effort $L_n/n$, and (h) shows the logarithm of the autocorrelation time in terms of computational effort $n\tau$.
    }
    \label{fig:ala_loss_diff_b}
\end{figure*}

We start by investigating the correlation between the loss and the autocorrelation time of the potential energy.
In Fig.~\ref{fig:ala_loss_diff_b}(a)-(c) we plot the loss surfaces for different choices of $b$ in the definition of the loss of Eq.~(\ref{eq:loss}).
A larger value of $b$ promotes big jumps and gives less importance to small jumps, that is, big moves are more important.
We have checked that the surface has multiple minima/maxima when we do no jittering, reiterating the importance of its inclusion.
In Fig.~\ref{fig:ala_loss_diff_b}(d), the corresponding autocorrelation time based on the potential energy is shown.
This approach of adapting the loss function is similar in spirit to the one presented in Ref.~\onlinecite{l2hmc}, although the influence was not investigated in detail there.
They chose to include a reciprocal term with a positive sign, i.e., actively small jumps in the coordinates were penalized.
We have empirically checked the loss proposed in this reference, but did not find suitable parameters for their free parameter which corresponded to a better match.
\par
We find that using the common definition of a squared jump distance ($b=2$) in Fig.~\ref{fig:ala_loss_diff_b}(a) does not correlate well with the actual observed autocorrelation times in Fig.~\ref{fig:ala_loss_diff_b}(d).
The region having a small loss is very large, going down to a small number of integrations steps $n \approx 3$ for $\Delta t=2.3$~fs, whereas for the autocorrelation times, the region of the minimum starts around $n\approx 9$.
As discussed before in Section~\ref{sec:obj}, there can be several reasons for this mismatch.
On the one hand, the limitation of optimizing for the lag-1 autocorrelation (made necessary to have a fast converging measure) is a potential source of mismatch and on the other hand, the focus on a different observable can introduce problems.
Thus, other definitions of the loss might provide a better correspondence.
\par
In (b) and (c) we therefore empirically test what happens to the loss surface for $b=3$ respectively $b=4$.
We find that for $b=3$, the minimum region of the loss shifts towards the right (larger number of integration steps $n$), as expected.
The correspondence between the loss surface and the autocorrelation times is much better.
For $b=4$, the minimum of the loss-surface appears still to align well with the autocorrelation time, however, it appears slightly too much favored towards large $n$.
This impression, however, changes once we consider the loss per computational cost $L_n/n$, which we plot in (e)-(g).
Due to the small differences in amplitude between the maximal and minimal loss values when compared to the differences between autocorrelation times, the division by $n$ massively shifts the loss towards smaller $n$, which is not reflected in the autocorrelation time.
As for the harmonic oscillator, we also here observe a less pronounced shift towards smaller $n$ for the autocorrelation time in units of computational effort $n\tau$ plotted in (h).
From these plots, we find the loss for $b=4$ in (g) to provide a good correspondence, so we will use this value for the following analysis.
We believe that this value may be well suited for a larger class of systems, although a physically more detailed study is necessary to make definitive statements.

\subsubsection{Learning of HMC Parameters}
\begin{figure*}
    \includegraphics{"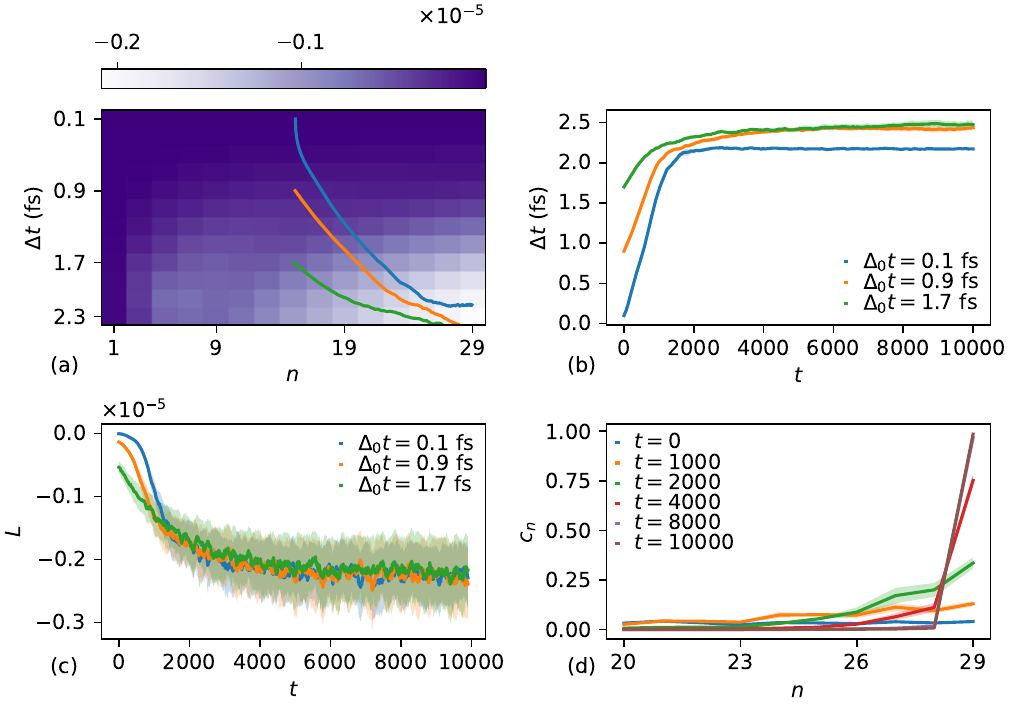"}
    \caption{(a) Loss per computational effort $L_n/n$ as a function of $\Delta t$ and $n$. Drawn are also three example trajectories obtained during training for initial $\Delta_0 t=0.1$~fs, $0.7$~fs, and $1.7$~fs.
    In (b) we show the value of $\Delta t$ as a function of the learning epoch $t$ for the three initial conditions.
    (c) Shows the corresponding loss for these three initial conditions as a function of training epoch $t$.
    (d) displays the weights $c_n$ as a function of the corresponding timestep $n$ for different training epochs $t$ as mentioned in the legend, where we have focused on the region of $n \geq 20$.
    In all cases (b)-(d), the shaded regions in the plot correspond to the error of the mean obtained by averaging over $5$ independent runs.}
    \label{fig:ala_learning}
\end{figure*}
We now turn to learning the parameters of HMC, following the general outline of the previous discussion for the harmonic oscillator.
The learning rate is set to $\epsilon=0.001$ and the other parameters of the Adam optimizer\cite{kingma2014adam} are kept at the default of PyTorch.
The epoch length is set to $10$ MC proposals.
All results shown in this section are averaged over $5$ independent learning trajectories for each initial $\Delta_0 t$, where each run was performed using a different random number seed responsible for the initialization of the weights $C_n$, the sampling of velocities, and the acceptance/reject step of HMC.
\par
In Fig.~\ref{fig:ala_learning}(a) we present again the loss surface per computational effort $L_n/n$ for $b=4$, where we have plotted three representative learning trajectories for three different initial values of $\Delta_0 t$, where the meaning of the color of the lines can be extracted from (b) and (c) of the same Figure.
The trajectories are obtained by initially setting $\Delta_0 t=0.1$~fs, $0.9$~fs, and $1.7$~fs respectively, whereas the $C_n$ are initialized randomly resulting in a mean value close to $N/2$.
We see that all simulations approach a very similar optimum of the loss, which also corresponds to a region where the autocorrelation times are small.
Note that the curves assume larger values of $\Delta t$ than the region for which we had originally performed our grid-search of parameters (15 values for $n$ and 12 values for $\Delta t$), as is also clear from Fig.~\ref{fig:ala_learning}(b) where we show $\Delta t$ as a function of learning epoch $t$.
The values of $\Delta t$ should be seen in the context of the ones used for classical MD simulations in the canonical ensemble.
There, to capture the fastest motions and to guarantee stable simulations, one typically uses a timestep of $0.5$~fs for this system (some approaches restrict the motion of Hydrogens, allowing via this trade-off a larger timestep).
The timestep $\Delta t\approx 2.5$ we find as optimal allows for a nearly five-fold faster simulation, although of course due to the acceptance/reject step of MC some trajectories are rejected.
In addition, MC guarantees the exact sampling of the true Hamiltonian since we have no effects due to the discretization of the timestep.
\par
It is interesting to note that the values of the acceptance rate around the optimal region are in the range from $50\%$ to $60\%$ and by this somewhat smaller than the predicted ideal value of $\approx 65\%$,\cite{beskos2013optimal} although still compatible.
\par
Fig.~\ref{fig:ala_learning}(c) shows the value of the loss $L$ as a function of training epoch $t$.
All simulations have similar behaving loss curves (with some differences during the initial training), which all arrive at very similar loss values.
This is because although the values of $\Delta t$ are quite different, they cannot be distinguished by the definition of the loss.
For the start with $\Delta_0 t=0.9$~fs, we plot also the weights given to each integration step for some selected epochs during training in  Fig.~\ref{fig:ala_learning}(d).
We find that starting from a random initialization giving every layer roughly the same weight, the weights move towards larger $n$ quite fast, resulting in a large weight for our maximally considered integration step $N=29$ for late training times $t$.
There is some interplay between the timestep $\Delta t$ and the number of integration steps $n$, as for a given (smaller) $\Delta t$ the optimum of $n$ does not need to coincide with the optimum for a different $n$ and thus the optimization due to the differential set-up shifts its respective optimum.
This, potentially, can influence the training, but we did not observe any obstacles in this regard.
\par
Our method thus allows the gradient-driven learning of good simulation parameters for HMC, saving a factor of above $100$ in computational effort ($5.4\times10^8$ force evaluations for the here performed grid-search vs. $2.9\times 10^6$ force evaluations for the gradient-based optimization).\footnote{The number of force evaluations are calculated as the product of the total number of epochs, the number of MC proposals per epoch, and the number of integration steps per MC proposal. In the case of the equilibrium runs, additionally we consider the number of different $\Delta t$ and $n$ we consider in the grid.}
This, of course, relies on a suitable definition of the loss as proposed here, which is certainly a limitation of adaptive MC methods in general.

\subsubsection{Atom Dependent Timesteps}
\label{sec:atom_dependent}
\begin{figure*}
    \includegraphics{"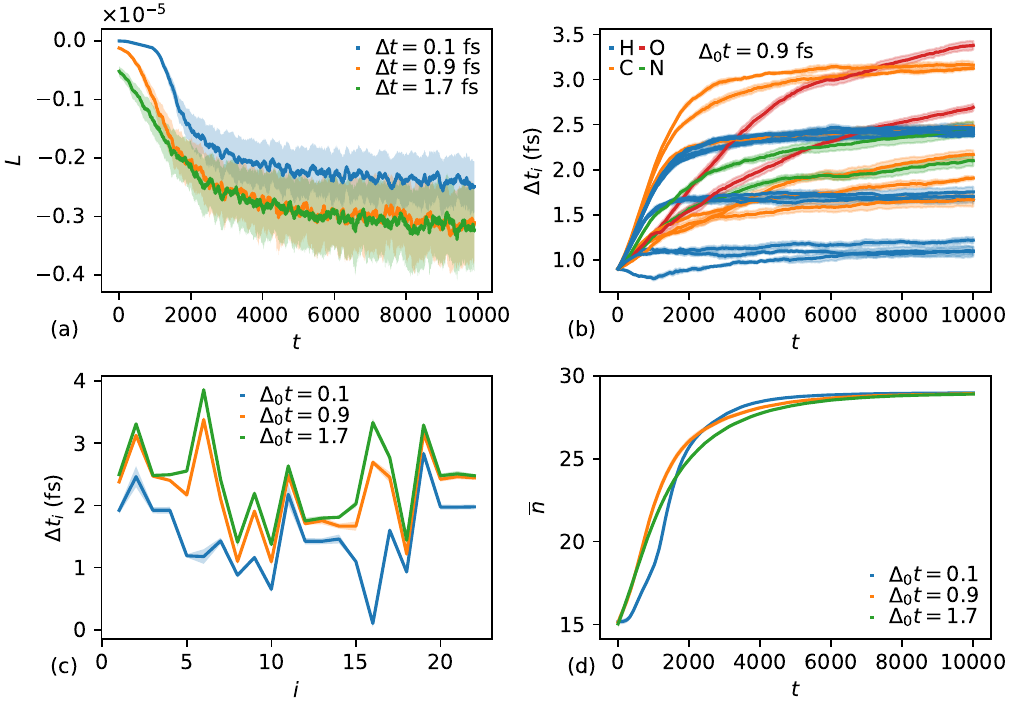"}
    \caption{(a) Loss $L$ as a function of epoch $t$ for different initial values of $\Delta_0 t$.
    (b) Shows the individual $\Delta t_i$ per atom for the initial value of $\Delta t=0.9$~fs, where we have colored atoms of the element with the same color.
    In (c) the final learned value of $\Delta t_i$ after training is shown for the different initial values of $\Delta t$.
    Finally, (d) shows the mean value of integration steps $\overline{n}$ as a function of learning epoch $t$.
        }
    \label{fig:ala_learning_dt_over_time}
\end{figure*}
\begin{table}
    \caption{Autocorrelation times $\tau$ for different initial $\Delta_0 t$ for atom based timesteps and global timesteps. In the brackets, we note the error of the mean.}
    \centering
    \begin{tabular}{c|c|c|c}
        \hline
        \hline
    $\Delta_0 t$ & $0.1$~fs & $0.9$~fs & $1.7$~fs\\
    \hline
    $\tau$ for atom based $\Delta t$ & $12.7(2.6)$ & $7.5(9)$ & $7.5(1)$ \\
    \hline
    $\tau$ for global $\Delta t$ & $12.1(1.8)$ & $10.0(1.0)$ & $9.9(1.3)$ \\
    \hline
    \hline
    \end{tabular}
    \label{tab:auto_vs_loss}
    \end{table}
    \begin{figure}
        \includegraphics{"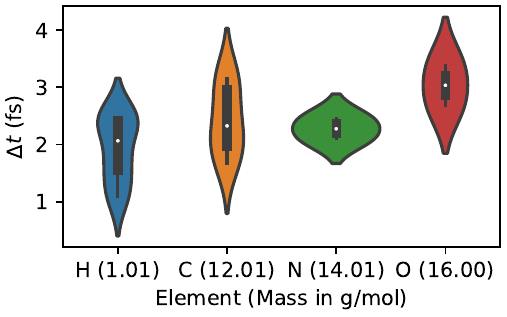"}
        \caption{Violin plot for the distribution of the timestep $\Delta t$ for the four different atom types present in alanine dipeptide, taken for the start with $\Delta_0 t=0.9$~fs. The elements are sorted according to their atom weight, which is mentioned in the brackets up to two digits.}
        \label{fig:dist_timestep}
    \end{figure}
In addition to learning the standard parameters of HMC, our approach also allows for the learning of many more parameters, such as e.g., atom-dependent timesteps.
This means, we now not only have a single $\Delta t$ for all atoms, but rather a different $\Delta t_i$ per atom index $i$.
As a reminder: We have $22$ atoms in alanine dipeptide which means that we now have $22$ different parameters for the timestep to optimize.
For the numbering of the atoms in the following Figures, refer to Fig.~\ref{fig:graph_ala} showing the molecule.
Heuristic approaches not based on gradients would most likely be much less efficient at the optimization of these many parameters, which highlights the importance of our fully differentiable approach.
\par
We use the same definition of the loss as in the last section, but since the parameter space is now of much higher dimension, the loss surface cannot any longer be explored by a grid search or even easily visualized, which is why we only consider improvements of the autocorrelation times directly.
We find that the overall optimization results in smaller loss values, as shown in Fig.~\ref{fig:ala_learning_dt_over_time}, as is expected for a more parameterized version of the integrator.
Compared to only having a single time-step $\Delta t$, we now find a loss value after training of about $L \approx -0.3\times10^{-5}$ (Fig.~\ref{fig:ala_learning_dt_over_time}(a)) compared to $L \approx -0.2\times10^{-5}$ (Fig.~\ref{fig:ala_learning}(c)), at least for starts with $\Delta_0 t=0.9$~fs and $1.7$~fs.
\par
In Table~\ref{tab:auto_vs_loss} we present the autocorrelation times for the potential energy.
We find that using atom-based $\Delta t_i$, our autocorrelation times for $\Delta_0 t=0.9$~fs and $1.7$~fs are roughly $25\%$ lower compared to their counterpart having a global $\Delta t$.
Using atom-based $\Delta t_i$ has only very little influence on the resulting wall-clock runtime after training, so incorporating them in practice simply results in the reported speed-up without additional cost.
For $0.1$~fs, the autocorrelation times using atom-based timesteps or a global timestep are comparable, which is not surprising since for this case the loss values are also comparable.
\par
In Fig.~\ref{fig:ala_learning_dt_over_time}(b) we plot the atom-based timesteps $\Delta t_i$ as a function of epoch $t$ for initial $\Delta_0 t=0.9$.
We find that the timestep for some atom index $i$ is greatly improved relative to others, which corresponds to a larger timestep of these atoms.
Also, the absolute value is much larger than the average value one obtains when optimizing only the ``global'' timestep $\Delta t$.
Figure \ref{fig:ala_learning_dt_over_time}(c) shows the final values of $\Delta t_i$ after the learning, which highlights that, at least for the initial values of $\Delta_0 t=0.9$~fs and $1.7$~fs, one arrives at very similar behavior of $\Delta t_i$, which is an indicator that there is a local minimum of the loss for these values of $\Delta t_i$.
There is up to a $3.5$ fold difference between the largest and smallest timestep, highlighting the differences in the ideal parameters.
For $\Delta_0 t=0.1$~fs, many signatures as for the other two initial starting parameters remain, i.e., for many atoms the values of $\Delta t_i$ follow the same trend as observed for $\Delta_0 t=0.9$~fs and $1.7$~fs, although it is unclear whether they would converge to the exactly same value in the long run using local gradient-based optimizers.
Most likely the optimization is trapped in a different local minimum, going back to our discussion about these potential limitations of our approach in Section~\ref{sec:jittering_local}.
For all initial values of $\Delta_0 t$, we find that for the number of integration steps $n$ we approach $n=29$, i.e., our currently maximal allowed number of integration steps, as shown in Fig.~\ref{fig:ala_learning_dt_over_time}(d).
\par
To gain some physical insight into the obtained values for $\Delta t_i$, we plot in Fig.~\ref{fig:dist_timestep} the distribution of them for the different atom types of alanine dipeptide, where the values of $\Delta t_i$ are obtained after training with $\Delta_0 t=0.9$~fs.
The elements are ordered according to their mass, showing some positive correlation between the mass of the atom and the ideal timestep $\Delta t_i$.
The distribution, however, is very broad in many cases.
This implies that this is indeed not a simple property of atom type only, but is most likely determined by its local neighborhood and the temperature.
For a more detailed analysis, significantly more data for different physical systems and temperatures would be needed, which we take as an interesting endeavor.

\section{Conclusion \& Outlook}
\label{sec:con}
We have presented a framework that allows for the gradient-based tuning of the simulation parameters of Hamiltonian Monte Carlo.
Its capabilities are evaluated for the one dimensional harmonic oscillator that provides crucial insights into the properties of our approach and alanine dipeptide as a more realistic test system.
The experiments show that, in both systems, our set-up allows for the optimization of the parameters of Hamiltonian Monte Carlo, leading to fast simulations and low values of the autocorrelation time without the need for an expensive grid search for ideal parameters.
Compared to a grid search, we observe a $>100$ fold speed-up for alanine dipeptide in obtaining good simulation parameters.
This success crucially depends on a local proxy loss for the autocorrelation time for which we propose a form with only one free parameter that works well for alanine dipeptide, with potentially more general character for the application in other molecular systems.
The definition of the loss and its generality is thus a critical ingredient to the gradient-driven optimization, making it a prime candidate for further investigations.
\par
We also show that jittering of the timestep avoids local minima in the loss surface, which is crucial for the optimization, which would otherwise get stuck in local basins of attraction and would not be able to find good values.
Further, enabled by the gradient-driven optimization approach, we extend the parameters of the integrator for alanine dipeptide by introducing timesteps that depend on the atom index.
We find that this can lead to lower loss values as compared to using a global timestep, which is also reflected in a $\approx 25\%$ lower autocorrelation time without additional computational overhead.
\par
Investigating the performance improvement potential for more complex systems is highly interesting.
An interesting use case of our algorithm are, for example dense, polymer melts.
The dynamics of these systems below the glass transition temperature is very slow, thus providing a challenging target system for simulation methods.\cite{kampmann2015monte,mavrantzas2021using}
It is unclear how much can be gained by the local optimization in this case, and whether a significant speed-up can be achieved.
Another direction to further speed up the simulations in this regard is the combination with hand-crafted Monte Carlo updates, where Hamiltonian Monte Carlo plays the role of a particular move.
In such a setting, the target is not only optimize the parameters of Hamiltonian Monte Carlo, but also the parameters of the distribution for the other moves and their relative pick probability.
\par
Also of interest are ``data-driven'' integrators, i.e., it would be interesting to investigate much more heavily parameterized versions of the integrator, for example by including (graph) neural networks into the integrator.\cite{bacciu2020gentle}
However here, in contrast to our method, the additional cost of evaluating the neural networks has to be considered, which can reduce some advantages of a parametrization with many parameters.

\begin{acknowledgments}
We thank Viktor Zaverkin, Makoto Takamoto, and Mathias Niepert for useful discussion.
\end{acknowledgments}

%
\end{document}